\documentstyle[11pt,epsf]{article}
\newcommand{\beq}{\begin{equation}}
\newcommand{\eeq}{\end{equation}}

\begin{document}

\title{The integrable dynamics of discrete and continuous curves\thanks{This
work was supported by KBN grant 2-0168-91-01, by the INFN and by the 1994
agreement between Warsaw and Rome Universities}}
\author{ Adam Doliwa\\
Institute of Theoretical Physics, Warsaw University \\
ul. Ho\.{z}a 69, 00-681 Warsaw, Poland \\
e-mail: doliwa@fuw.edu.pl \\
\and
Paolo  Maria Santini  \\
Dipartimento di Fisica, Universit\`{a} di Catania \\
Corso Italia 57, I-95129 Catania, Italy \\
and INFN, Sezione di Roma, P.le Aldo Moro 2, I-00185 Roma, Italy \\
e-mail: santini@catania.infn.it  \/  santini@roma1.infn.it }
\date{}
\maketitle

\begin{abstract}
\noindent We show that the following geometric properties of the motion of
discrete and
continuous curves select integrable dynamics: i) the motion of the curve takes
place in the $N$ dimensional sphere of radius $R$, ii) the curve does not
stretch during the motion, iii) the equations of the dynamics do not depend
explicitly on the radius of the sphere. Well known examples of integrable
evolution equations, like the nonlinear Schr\"{o}dinger and the sine-Gordon
equations, as well as their discrete analogues, are derived in this general
framework.
\end{abstract}
\section{A historical introduction}
\label{sec:history}
One of the classical problems of the XIX-century geometers was the
study of the connection between differential geometry of submanifolds and
nonlinear (integrable) PDE's. For instance, Liouville found the general
solution of the equation (known now as the Liouville equation) which describes
minimal surfaces in $E^{3}$~\cite{Liouville}.
Bianchi solved the general Goursat problem for the sine-Gordon (SG)
equation~\cite{SG}, which encodes the whole geometry of the pseudospherical
surfaces. Moreover the method of construction of a  new pseudospherical surface
from a given one, proposed by Bianchi~\cite{Bianchi},
gives rise to the B\"{a}cklund transformation for the SG
equation~\cite{Backlund}.

The connection between geometry and integrable PDE's
became even deeper when Hasimoto~\cite{Hasimoto} found the transformation
between the equations governing the curvature and torsion of a nonstretching
thin vortex filament moving in an incompressible  fluid and
the NLS equation. Several authors, including Lamb \cite{Lamb}, Lakshmanan
\cite{Lakshmanan}, Sasaki \cite{Sasaki}, Chern and Tenenblat
\cite{Chern and Tenenblat}
related the Zakharov-Shabat(ZS) \cite{Z-S} spectral problem and the associated
Ablowitz-Kaup-Newell-Segur(AKNS) hierarchy \cite{AKNS} to the motion of curves
in $E^{3}$ or to the pseudospherical surfaces and certain foliations on them.

Almost at that time Sym introduced the soliton surfaces
approach, in which the powerful tools of the IST method are
used to construct explicit formulas for the immersions of one-parameter
families (labeled by the spectral parameter) of surfaces corresponding to
given solutions of integrable PDE's~\cite{Sym}; see also the recent
developments of Bobenko~\cite{Bobenko}.

More recently Langer and Perline~\cite{Langer-Perline} showed that the
dynamics of a nonstretching vortex filament in $R^{3}$ gives rise, through the
Hasimoto transformation, to the recursion operator  of the NLS hierarchy.
Similarly, Goldstein and Petrich~\cite{GolPe} showed that the
dynamics of a nonstretching string on the plane produces the recursion operator
of the mKdV hierarchy.

Connections between geometry and integrable PDE's in multidimensions can also
be found, for example in works of Tenenblat and Terng~\cite{TenTer} and
Konopelchenko \cite{Konop}.
Also at a discrete level there is a similar situation. For instance, discrete
pseudospherical surfaces and the
discrete analogues of constant mean curvature surfaces are described by
integrable discrete analogues of the sine-Gordon (SG) \cite{Wun}\cite{BoPin}
and sinh-Gordon
equations \cite{Pin}. Such discretizations were found by adapting the main
geometric properties of the continuous surfaces to a discrete level.

In two recent papers \cite{DolSan1}\cite{DolSan2} we have proposed a new
geometric  characterization of the integrable dynamics of a discrete or
continuous  curve (where, by discrete curve, we mean just a sequence of
points), based on the following three properties:

\medskip

\noindent {\it Property 1.} The motion of the curve takes place in
 the $N$-dimensional sphere of radius $R$, denoted
by $S^{N}(R)$, $N>1$.

\medskip
\noindent {\it Property 2.} The curve does not stretch during the motion.

\medskip
\noindent {\it Property 3.} The equations of the dynamics of the curve do not
depend {\bf explicitly} on the radius $R$.

We remark that {\it Properties 1-3} not only select
integrable PDE's, but also provide their integrability scheme; in other words,
in the process of deriving the dynamics selected by {\it Properties 1-3} one
discovers "for free" the integrable nature of such dynamics! In particular, the
spectral problem is given by the Frenet equations of the curve and is a
consequence of {\it Property 1}, and the spectral parameter is given by the
inverse of the radius of the sphere.

We also remark that our aproach explains in a simple way Sym's formula
\cite{Sym}, which allows to calculate, from  the wave function of the spectral
problem, the surface generated by the motion of the curve.

In papers \cite{DolSan1} and \cite{DolSan2} we have dealt with the integrable
dynamics of a continuous and discrete curve respectively, obtaining, in the
case of $N=3$, the AKNS \cite{AKNS} and the Ablowitz-Ladik (AL)  \cite{AL}
hierarchies. As we shall see in the following if, in {\it Property 3}, we
consider discrete time dynamics, one also generates integrable fully discrete
evolution equations, like the Hirota equation \cite{Hir}\cite{Orf}.
Since integrable discrete dynamics can always be interpreted as
B\"{a}cklund Transformations (BT's) of the corresponding continuous dynamics
\cite{LeBen}, the hierarchies of BT's of integrable systems are also
characterized by {\it Properties 1--3}.

\section{The curve in $S^{N}(R)$ and the associated spectral problem}
\subsection{Frenet basis along the discrete curve}
Let us consider a sequence ${\bf Z} \ni k \mapsto {\bf r}(k) \in S^{N}(R)
\subset {\bf R}^{N+1}$ of points of the N-dimensional sphere of radius $R$.
We are interested in the sequence ${\bf r}(k)$ which gives rise to a piecewise
linear curve in $S^{N}(R)$. Our goal is to construct an analog of the
Frenet basis along the discrete curve and of the corresponding Frenet
equations.

By $F^{1}(k)$ we denote the 1-dimensional oriented vector subspace of ${\bf
R}^{N+1}$ given by ${\bf r}(k)$ and let $F^{l+1}(k) = F^{l}(k) + F^{1}(k+l)$.
For the point ${\bf r}(k)$ in general position (what we assume in the sequel
for simplicity) $F^{l}(k)$ is $l$-dimensional
oriented subspace of ${\bf R}^{N+1}$ $(l\leq N+1)$. It is also convenient to
denote
$F^{0}(k) =\{ 0 \}$. Now we define the orthonormal Frenet basis
$\{ {\bf f}_{l}(k) \}_{l=0}^{N}$ in the point ${\bf r}(k)$ of the discrete
curve: ${\bf f}_{l}(k)$ is the unit vector of $F^{l+1}(k)$ orthogonal to
$F^{l}(k)$ and correctly oriented.

The distance $\Delta(k)$ between points ${\bf r}(k)$ and ${\bf r}(k+1)$ of the
curve is given in terms of the radius $R$ of the sphere and the angle
$\varphi_{0}(k)$ between ${\bf f}_{1}(k)$ and ${\bf f}_{1}(k+1)$ as
\beq \Delta(k) = R\, \varphi_{0}(k) \; \; . \eeq
We are going to construct $N-1$ other angles which play similar role as
curvatures of the (continuous) curve.

Both $F^{2}(k)$ and $F^{2}(k+1)$ are subspaces of $F^{3}(k)$
and their intersection is $F^{1}(k+1)$. Its orthogonal complement in $F^{3}(k)$
is the plane $\pi_{1}(k)$. Since ${\bf f}_{2}(k) \in F^{3}(k)$ and
${\bf f}_{2}(k)
\bot F^{2}(k)$, then ${\bf f}_{2}(k) \in \pi_{1}(k)$; similarly ${\bf
f}_{1}(k+1) \in \pi_{1}(k)$. By $\tilde{\bf f}_{1}(k)$ we denote the unit
vector of  $\pi_{1}(k)$ normal to ${\bf f}_{2}(k)$; it is also the vector of
$\pi_{0}(k) =F^{2}(k)$ orthogonal to ${\bf f}_{0}(k+1)$. The angle
$\varphi_{1}(k)$ between the hyperplanes $F^{2}(k)$ and $F^{2}(k+1)$ in
$F^{3}(k)$ (equivalently, between $\tilde{\bf f}_{1}(k)$ and ${\bf
f}_{1}(k+1)$) is the angle of geodesic curvature.

\epsffile{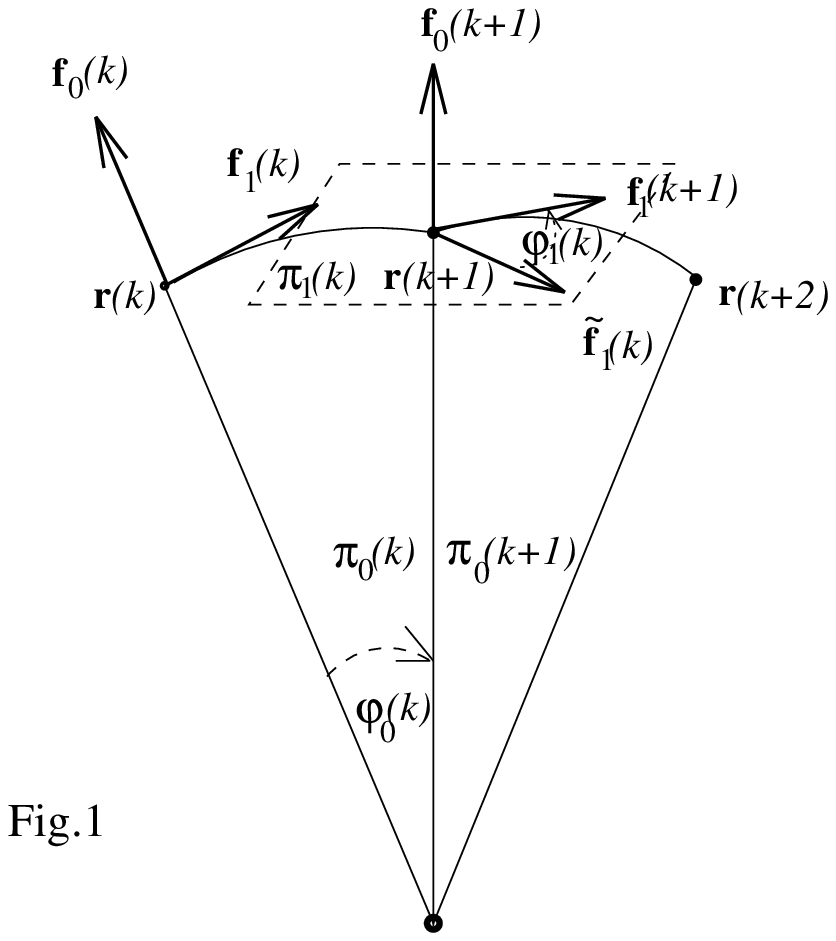}

\noindent In general we consider the subspaces $F^{l}(k)$ and $F^{l}(k+1)$ of
$F^{l+1}(k)$, their
intersection $F^{l-1}(k+1)$ and its orthogonal complement $\pi_{l-1}(k)$. Since
${\bf f}_{l}(k) \in F^{l+1}(k)$ and ${\bf f}_{l}(k) \bot F^{l}(k)$, then ${\bf
f}_{l}(k) \in \pi_{l-1}(k)$. Similarly, since ${\bf f}_{l-1}(k+1) \in
F^{l}(k+1)\subset F^{l+1}(k)$ and ${\bf f}_{l-1}(k+1) \bot F^{l-1}(k+1)$, then
${\bf f}_{l-1}(k+1) \in \pi_{l-1}(k)$.

By $\tilde{\bf f}_{l-1}(k)$ we denote the unit vector of $\pi_{l-1}(k)$
orthogonal to ${\bf f}_{l}(k)$. One can show that $\tilde{\bf
f}_{l-1}(k)$ is also the unit vector of $\pi_{l-2}(k)$ orthogonal to ${\bf
f}_{l-2}(k+1)$. This is the consequence of two facts: $\tilde{\bf f}_{l-1}(k)
\in F^{l}(k)$ (as $\tilde{\bf f}_{l-1}(k)\in F^{l+1}(k)$ and $\tilde{\bf
f}_{l-1}(k)\bot {\bf f}_{l}(k)$) and $\tilde{\bf f}_{l-1}(k) \bot
F^{l-1}(k+1)$.  The angle
$\varphi_{l-1}(k)$ between the hyperplanes $F^{l}(k)$ and $F^{l}(k+1)$ of
$F^{l+1}(k)$ (equivalently, between their normals ${\bf f}_{l}(k)$ and
$\tilde{\bf f}_{l}(k)$, or between $\tilde{\bf f}_{l-1}(k)$ and
${\bf f}_{l-1}(k+1)$) is the  angle of the $(l-1)$th curvature.

\epsffile{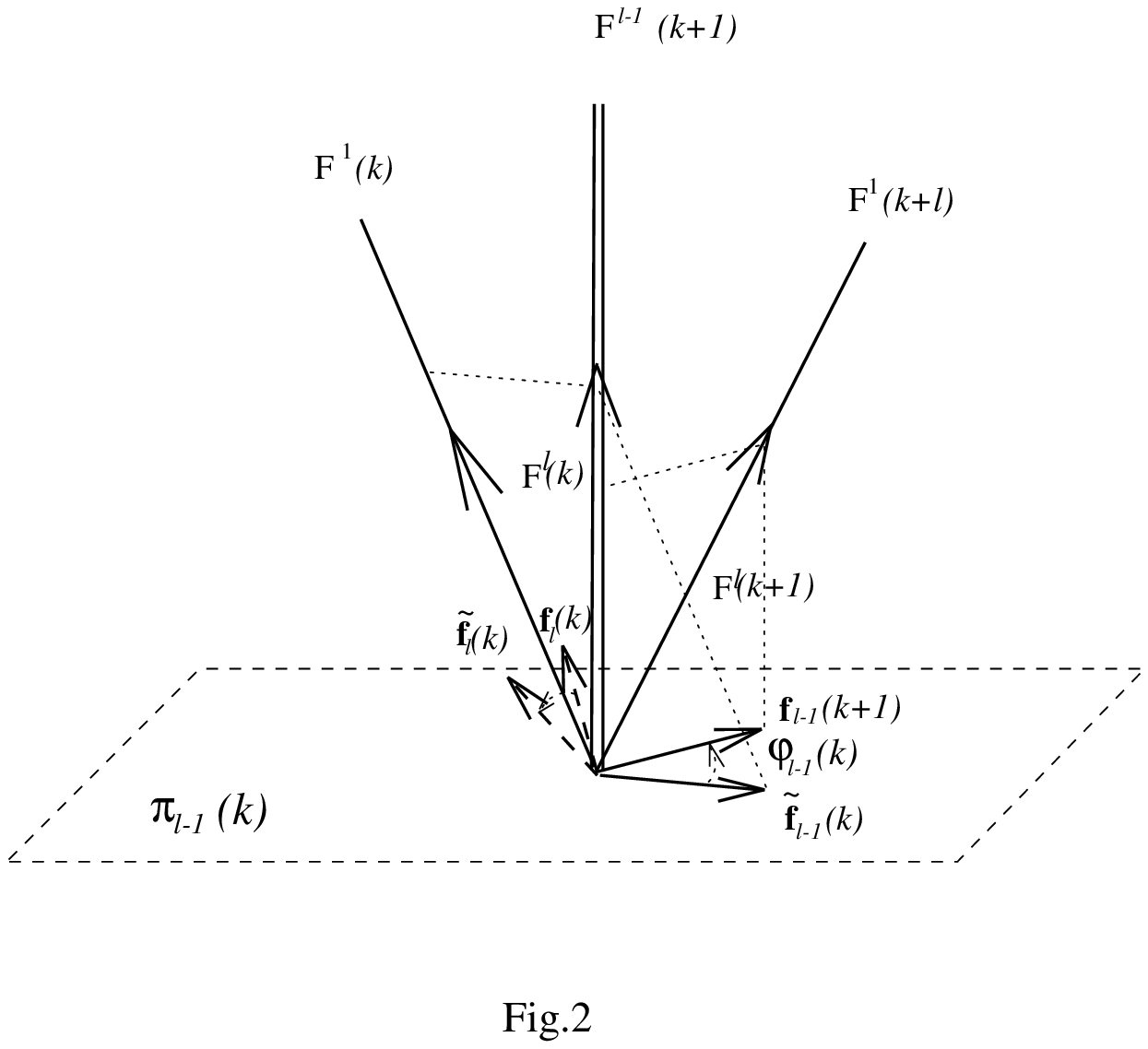}

\noindent The transition from the Frenet basis
$\{ {\bf f}_{l}(k) \}_{l=0}^{N}$ in the
point ${\bf r}(k)$ of the discrete curve to $\{ {\bf f}_{l}(k+1) \}_{l=0}^{N}$
is obtained by the superposition of N rotations of the angles
$\varphi_{l}(k)$ in the planes $\pi_{l}(k)$, $(l=0,...,N-1)$:
\[
\{{\bf f}_{0}(k),{\bf f}_{1}(k),{\bf f}_{2}(k),...,{\bf f}_{N}(k)\}
\stackrel{\varphi_{0}(k),\pi_{0}(k)}{\longrightarrow} \]
\[ \{{\bf f}_{0}(k+1),\tilde{\bf f}_{1}(k),{\bf f}_{2}(k),...,{\bf f}_{N}(k)\}
\stackrel{\varphi_{1}(k),\pi_{1}(k)}{\longrightarrow} ... \]
\beq ...\; \; \{{\bf f}_{0}(k+1),...,\tilde{\bf f}_{l-1}(k),
{\bf f}_{l}(k),{\bf f}_{l+1}(k),...,{\bf f}_{N}(k) \}
\stackrel{\varphi_{l-1}(k),\pi_{l-1}(k)}{\longrightarrow} \eeq
\[ \{{\bf f}_{0}(k+1),...,{\bf f}_{l-1}(k+1),\tilde{\bf f}_{l}(k),{\bf
f}_{l+1}(k),...,{\bf f}_{N}(k) \}
\stackrel{\varphi_{l}(k),\pi_{l}(k)}{\longrightarrow} ... \]
\[ ... \; \;  \{{\bf f}_{0}(k+1),...,{\bf f}_{N-1}(k+1),
\tilde{\bf f}_{N}(k)={\bf f}_{N}(k+1) \}
\; \; . \]

\subsection{The spinor representation of the Frenet equations}
The linear transformation related to the resulting rotation gives the discrete
analog of the Frenet equations. We will present it using the language of the
Clifford algebras and Spin groups \cite{BuTr}.
Let ${\cal E} = \{ {\bf e}_{l} \}_{l=0}^{N}$ be a fixed orthonormal basis of
${\bf R}^{N+1}$ cosidered as a subspace of the Clifford algebra
${\rm Cl}(N+1)$,
then  any other orthonormal (correctly oriented) basis ${\cal F}= \{ {\bf
f}_{l} \}_{l=0}^{N} $ can be
obtained from ${\cal E}$ using an element $S$ of the corresponding ${\rm
Spin}(N+1)$ group
\beq {\cal F} = S^{-1} {\cal E} S \; \; . \eeq
Moreover, when another basis ${\tilde {\cal F}}$ is obtained from
${\cal F}$ by rotation in
the plane $\langle{\bf f}_{i},{\bf f}_{j}\rangle$ of the angle $\varphi$,
then
\beq \tilde{S} = (\cos\frac{\varphi}{2} +
{\bf e}_{i}{\bf e}_{j}\sin\frac{\varphi}{2})S
 = O_{ij}^{\varphi} S\; \; . \eeq
If $S(k)\in {\rm Spin}(N+1)$ represents the rotation to Frenet basis in
point ${\bf r}(k)$ then it is subjected to the equation
\beq
\label{eq:FrenetD}
S(k+1) = O_{N-1,N}^{\varphi_{N-1}(k)}\cdot \dots \cdot
O_{12}^{\varphi_{1}(k)}O_{01}^{\varphi_{0}(k)}  S(k) \; \; .
\eeq
The arc-length along the curve
\beq
s= \sum_{i}^{k-1}\Delta(i) \; \;
\eeq
in the continuous limit $\varphi_{0}(i) \rightarrow 0$ is the arc-length
parameter. Moreover
\beq \frac{dS(s)}{ds} = \lim_{\varphi_{0}(k)\rightarrow 0}\frac{S(k+1) - S(k)}
{\Delta(k)} = \left( \frac{1}{R} E_{01} + \kappa_{1}(s)E_{12} + ... +
\kappa_{N-1}(s)E_{N-1,N} \right) S(s) \; \; ,
\label {eq:FrenetS} \eeq
where
\beq
\kappa_{l}(s)=\lim_{\varphi_{0}(k)\rightarrow
0}\frac{\varphi_{l}(k)}{\Delta(k)} \eeq
is the $l$-th curvature of the corresponding continuous curve and
$E_{ij} = {\bf e}_{i}{\bf e}_{j}/2$ are elements of the canonical basis of the
orthogonal Lie algebra so$(N+1)$ in the Clifford algebra representation. The
equation (\ref{eq:FrenetS}) is nothing but the classical Frenet equation for
the curve in $S^{N}(R)$.

\bigskip

\noindent {\bf Remark:} Even when point ${\bf r}(k)$ is not in general
position, the
corresponding spaces $F^{l}(k)$ can be defined in a way that their
dimension is $l$. We just keep the space $F^{l}(k-p)$ ($p>0$) of the nearest
point in which it was "properly" defined.
In this way one can define, for example, the Frenet frame along the geodesic
line
(which is any big circle): only ${\bf f}_{0}$ and ${\bf f}_{1}$ vary along the
curve, and the
rest of the Frenet frame remains as it was defined in a starting
point.

\section{The discrete curve in $S^{3}(R)$}
In this Section we investigate in detail the discrete curve in $S^{3}(R)$ with
constant distance $\Delta$ between the subsequent points. Throught the Section
we use the following definitions: $ \nu = \Delta/R = \varphi_{0}(k), \;
\varphi(k) = \varphi_{1}(k), \;  \theta(k)
= \varphi_{2}(k)$, the Frenet basis $\{ {\bf f}_{l}(k) \}_{l=0}^{3}$ is denoted
by $\{ {\hat {\bf r}}(k), {\bf t}(k), {\bf n}(k), {\bf b}(k) \}$ and consists
of the radial, tangent, normal and binormal vectors.

We show that the Frenet equation (\ref{eq:FrenetD}) reduces to the
Ablowitz-Ladik spectral problem, and
its continuous limit to the Zakharov-Shabat spectral problem. We also present
the geometric explanation of Sym's formula.

\subsection{The Hasimoto transformation for the discrete curve in $S^{3}$ and
the Ablowitz-Ladik spectral problem}

It is convenient to modify Frenet basis by a rotation in the normal plane
$\langle {\bf n}(k), {\bf b}(k)\rangle$ of the angle
$\sigma (k) = \sum_{i}^{k-1}\theta(i)$.
\beq
{\bf N}(k) = \cos \sigma(k){\bf n}(k) - \sin \sigma(k) {\bf b}(k)
\eeq
\[ {\bf N}_{J}(k) = \sin \sigma(k) {\bf n}(k) + \cos \sigma(k) {\bf b}(k)
\; \; .\]
This change of basis corresponds to a partial "integration" of the Frenet
equations in the normal plane, since the vectors ${\bf N}(k),{\bf N}_{J}(k)$
does not vary from the point of view of the normal plane (this is the discrete
analog of the parallel transport in the normal bundle).

It is also convenient to interprete any vector of the normal plane  as a
complex number
\beq
{\bf \vec{\phi}}(k) ={\rm Re}\phi(k){\bf N}(k) +{\rm Im}\phi(k) {\bf N}_{J}(k)
 \; \; \Leftrightarrow \; \; \phi(k) = {\rm Re} \phi(k) + i\:
{\rm Im} \phi(k) \; \; .
\eeq
If we define $S(k) \in {\rm Spin(4)}$ by the relation
\[ {\cal H}(k)=\{ {\hat {\bf r}}(k), {\bf t}(k), {\bf N}(k), {\bf
N}_{J}(k) \} = S(k)^{-1} {\cal E} S(k)\; \; , \]
then
\[
S(k+1) = O_{23}^{-\sigma(k)}O_{12}^{\varphi(k)}O_{23}^{\sigma(k)}
O_{01}^{\nu} S(k) \; \; . \]
To represent the above rotation in terms of matrices we first choose the
following representation of the basis ${\cal E}$ as $4\times 4$ Dirac matrices
\beq
{\bf e}_{0} \leftrightarrow \left( \begin{array}{cc}
0 & {\rm I} \\
{\rm I} & 0  \end{array} \right) \; \; \; , \; \; \;
{\bf e}_{1} \leftrightarrow \left( \begin{array}{cc}
0 & -i\sigma_{3} \\
i\sigma_{3} & 0  \end{array} \right) \; \; \; , \eeq
\[{\bf e}_{2} \leftrightarrow \left( \begin{array}{cc}
0 & -i\sigma_{1} \\
i\sigma_{1} & 0  \end{array} \right) \; \; \; ,\; \;
{\bf e}_{3} \leftrightarrow \left( \begin{array}{cc}
0 & i\sigma_{2} \\
-i\sigma_{2} & 0  \end{array} \right) \; \; ,
\]
where  I is the $2\times 2$ identity matrix and
$\sigma_{l}$ are the standard Pauli matrices .
The linear problem related to the discrete curve takes the form
\beq \label{eq:S}
S(k+1) = \left( \begin{array}{cc} S'(k+1) & 0 \\
0 & S''(k+1) \end{array} \right) =
\left( \begin{array}{cc} A'(k) & 0 \\
0 & A''(k) \end{array} \right) \left( \begin{array}{cc} S'(k) & 0 \\
0 & S''(k) \end{array} \right) \; \; , \eeq
with
\[
A'(k)= \left( \begin{array}{cc}
{\rm e}^{i\nu/2}\cos(\varphi(k)/2)  &
{\rm e}^{-i\nu/2}\sin(\varphi(k)/2) {\rm e}^{i\sigma(k)} \\
-{\rm e}^{i\nu/2}\sin(\varphi(k)/2){\rm e}^{-i\sigma(k)} &
{\rm e}^{-i\nu/2}\cos(\varphi(k)/2)
\end{array} \right) = \]
\beq  =  \frac{1}{\sqrt{1+|q(k)|^{2}}} \left( \begin{array}{cc}
\zeta  & q(k)\zeta^{-1}   \\
-\bar{q}(k)\zeta & \zeta^{-1} \end{array} \right)   \; \; , \eeq
wher
\beq
q(k)=\tan (\varphi(k)/2){\rm e}^{i\sigma(k)} \; \; , \; \; \zeta={\rm
e}^{i\nu/2} \; \; ,
\eeq
and $A''(k,\zeta) = A'(k,\zeta^{-1})$. This linear
problem is equivalent \cite{DolSan2} to the Ablowitz-Ladik spectral problem
\cite{AL}.

In the limit of the continuous curve we obtain the Zakharov-Shabat spectral
problem \cite{Z-S}
\beq
\frac{d S'(s)}{ds} = \frac{1}{2}\left( \begin{array}{cc} i\lambda & q(s) \\
-\bar{q}(s) & -i\lambda \end{array} \right) S'(s) \; \; ,
\eeq
where $q(s) = \kappa(s) {\rm e}^{i\sigma(s)}$, $\sigma(s) =
\int^{s}\tau(s')ds'$ and $\lambda = R^{-1}$.

\subsection{The geometric interpretation of Sym's formula}
In short notation: ${\cal E}\leftrightarrow \{ {\rm I}, i\sigma_{3},
i\sigma_{1}, -i\sigma_{2} \}$
\beq {\cal H}(k) = S''(k)^{-1}{\cal E} S'(k) \eeq
and, as a consequence, for a function $q(k)$, the radius vector of the
corresponding curve in sphere $S^{3}(R)$ is represented by
\beq
{\bf r}(k) = R \, S''(k)^{-1}\,  S'(k) \; \; .
\eeq
Suppose one is interested in the radius vector of the curve  in ${\bf R}^{3}$
corresponding to $q(k)$. One can consider ${\bf R}^{3}$ as sphere of infinit
radius but one cannot just take the limit $R\rightarrow \infty$ in the formula
above. This way the center of the sphere is fixed while ${\bf R}^{3}$ is pushed
away to infinity.

\epsffile{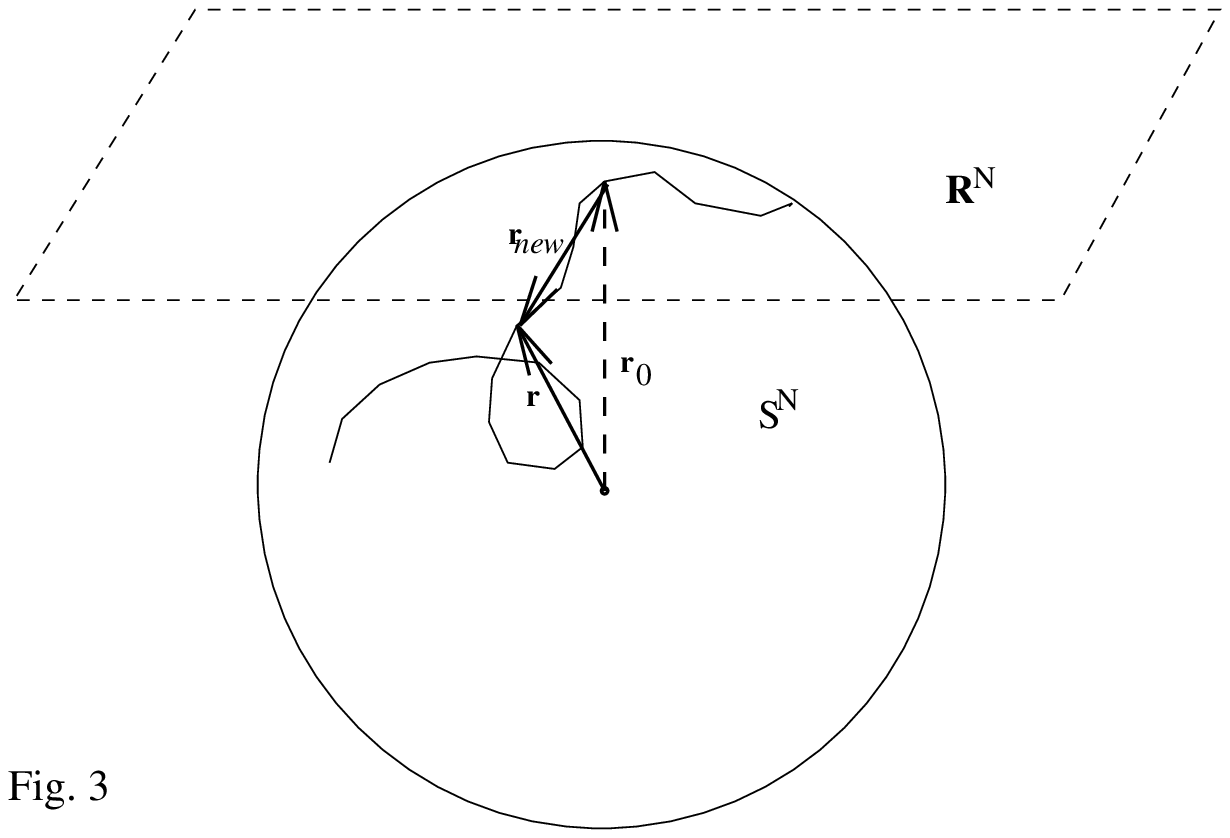}

\noindent To remove this incovenience we first have to shift the basis of ${\bf
R}^{4}$  to a point of the sphere: ${\bf r}_{new}(k)  = {\bf r}(k) -{\bf
r}_{0}$.

If the linear problem is solved under the initial condition
$S'(0,\lambda)= S''(0,\lambda) = {\rm I}$, then $S''(k,\lambda) =
S'(k,-\lambda)$. Choosing  ${\bf r}_{0} = {\bf r}(0,\lambda) = R {\bf e}_{0}
= \frac{1}{\lambda} {\rm I}$, one obtains the following formula for the
Cartesian
coordinates $(X^{i}(k))_{i=1}^{3}$ of the points $\tilde{\bf r}(k)$ of the
curve in ${\bf R}^{3}$
\beq
\tilde{\bf r}(k) = \sum_{i=1}^{3}X^{i}(k){\bf e}_{i} =
\lim_{\lambda\rightarrow 0} \frac{1}{\lambda}
\left( S'(k,\lambda)^{-1}S'(k,\lambda) - {\rm I} \right) =
2 S'(k,0)^{-1} \frac{\partial S'(k,\lambda)}{\partial \lambda}|_{\lambda
= 0} \; \; . \eeq
The above formula was first used by Sym \cite{Sym} in his approach of soliton
surfaces. He also found its generalization to soliton equations related to
linear problems in semi-simple Lie algebras. Our approach is more
related to the Clifford algebras.

\medskip
\noindent {\bf Remark:} A modification of this formula
was recently found by Cie\'{s}li\'{n}ski \cite{Cieslinski}
in the context of conformal geometry of isothermic surfaces in ${\bf R}^{3}$.

\section{The integrable dynamics in $S^{3}(R)$}
In this Section we consider the motion of the discrete curve subjected to {\it
Properties 1 - 3}. For convenience, we use the following short-hand notation:
$f$ for $f(k), k\in {\bf Z}$ and $f_{n}$ for $f(k+n), \; n=\pm 1,\pm  2,..$.

The motion of the curve is governed by velocity field ${\bf v}$ which is
convenient to write in the form
\beq \label{eq:kin_r}
{\bf r}_{,t} = {\bf v} =  \frac{\sin(\lambda\Delta)}{\lambda} \left( V {\bf t}
+ {\rm Re}\phi{\bf N} +{\rm Im}\phi {\bf N}_{J}  \right) \; \; ,
\eeq
and, consequently,
\beq \label{eq:evS}
S'_{,t} = T S'= \frac{i}{2}\left( \begin{array}{cc}
\gamma & \delta \\
\bar{\delta} & -\gamma \end{array} \right) S' \; \; \; , \; \; T\in {\rm su}(2)
\; \; .
\eeq
The compatibility condition between equations (\ref{eq:S})(\ref{eq:kin_r}) and
(\ref{eq:evS}) specifies the entries $\gamma$ and $\delta$ in terms of
the velocity field:
\begin{eqnarray}
\delta & = & i\zeta^{-2}\phi - i ( \phi_{1} + q (V + V_{1})) \nonumber \\
\gamma & = & W + \sin(\lambda\Delta) V \\
W_{1}- W & = & - {\rm Re} \left( i\bar{q} \left( \phi_{2} -\phi
+ q_{1} (V_{1} +V_{2}) \right) \right) \nonumber
\end{eqnarray}
and yields the kinematics
\beq \label{eq:kinAL}
2q_{,t}  =  -2\cos(\lambda\Delta)\left( \phi_{1} + qV_{1} \right) +
{\cal R}\left( \phi_{1} + qV_{1} \right)  \; ,
\eeq
\beq
(1-|q^{2}|) V_{1} - (1+|q|^{2}) V = q \bar{\phi}_{1} +
\bar{q} \phi_{1} \; \; ,
\eeq
where
\beq
{\cal R}f :=  (1+|q|^{2}) \left( f_{1} + f_{-1} + 2(q_{1}E -
q_{-1} ) (E-1)^{-1} {\rm Re}\frac{\bar{q}f}{1+|q|^{2}} \right)
- 2iq(E-1)^{-1}{\rm Im}(\bar{q}_{1} f - \bar{q} f_{1}) \; \;
\eeq
and $E$ is the shift operator along the discrete curve: $E f = f_{1}$.

Substituting the ansatz
\beq \label{eq:ansatz}
\left( \begin{array}{c} V \\ \phi \end{array} \right) = \sum_{j=0}^{m}
(\cos(\lambda\Delta))^{j} \left( \begin{array}{c} V^{(m-j)} \\
\phi^{(m-j)} \end{array} \right)
\eeq
into equation (\ref{eq:kinAL}) and requiring independence of
$\cos(\lambda\Delta)$, we finally obtain the following class of integrable
dynamics:
\beq
q_{,t} = h_{0}({\cal R}) (1+|q|^{2}) (q_{1} - q_{-1}) + h_{1}({\cal
R})(iq) \; \; ,
\eeq
where $h_{0}$ and $h_{1}$ are arbitrary entire functions with real
coefficients.

We remark that equation (\ref{eq:kinAL}) implies the following
interesting connection:
\beq
K^{(m)} = \phi_{1}^{(m)} + qV_{1}^{(m)}
\eeq
between the integrable commuting flows
\beq
K^{(m)} = {\cal R}^{m-1}(1+|q|^{2})(q_{1} - q_{-1}) \; \;
{\rm and/or} \; \; K^{(m)} = {\cal R}^{m}(iq) \; \; , \; \; m\geq 0
\eeq
and the velocity fields. In the continuous limit this reduces to the result of
Langer and Perline \cite{Langer-Perline}.

The simplest examples are the following:

\medskip

\noindent i) If $h_{0}=1$, $h_{1}=0$, then
\beq
{\bf v} = \frac{\sin(\lambda\Delta)}{\lambda}\left( {\bf t} -  \vec{q}_{-1}
\right) =
\frac{\sin(\lambda\Delta)}{\lambda}\left( {\bf t} - |q_{-1}| {\bf n}\right) \;
\; ,
\eeq
\beq
q_{,t} = (1+|q|^{2}) (q_{1} - q_{-1}) =: K^{(1)} \; \; .
\eeq
ii) If $h_{0}(x)=x$, $h_{1}=0$, then
\beq
{\bf v} = \frac{4\sin(\lambda\Delta)}{\lambda} \left( \left(
\cos(\lambda\Delta) + \frac{1}{2}(q\bar{q}_{-1} + \bar{q} q_{-1})
\right) {\bf t} +  \vec{\phi} \right) \; \; ,
\eeq
\[ \phi= \left( -q_{-1}\cos(\lambda\Delta) + \frac{1}{2}\left( (1+
|q_{-1}|^{2} )(q - q_{-2}) - q_{-1}(q\bar{q}_{-1} + \bar{q}
q_{-1} ) \right) \right) \]
\beq
q_{,t}=
( 1 + |q|^{2}) \left( ( 1 + |q_{1}|^{2} ) q_{2} - ( 1 +|q_{1}|^{2} )
q_{-2} +  \bar{q} ( q_{1}^{2} - q_{-1}^{2} ) + q ( q_{1}\bar{q}_{-1} -
q_{-1}\bar{q}_{1} ) \right)=:  K^{(2)} \; \; .
\eeq 
iii) If $h_{0}=0$ and $h_{1}(x)=x$, we obtain
\beq
{\bf v} = \frac{2\sin(\lambda\Delta)}{\lambda} \left( i\vec{q}_{-1}\right)
= \frac{2\sin(\lambda\Delta)}{\lambda} \, \tan(\frac{\varphi_{-1}}{2}) {\bf
b} \; \; ,
\eeq
\beq \label{eq:dNLS}
q_{,t} = i(1+ |q|^{2}) (q_{1} + q_{-1}) \; \; .
\eeq
This equation has also recently appeared in conection with the Heisenberg XXO
antiferromagnet model \cite{IIKS}.

In the continuous limit, ${\cal R}-2$ reduces to the recursion operator of the
continuous NLS hierarchy. Moreover
the following combination of equation (\ref{eq:dNLS}) with the "zero order
flow" $q_{,t}= i q$ :
\beq \label{eq:dNLS2}
q_{,t} = i\left( q_{1} - 2q + q_{-1} + |q|^{2}(q_{1} + q_{-1}) \right)
\eeq
reduces \cite{AL} in the continuous limit, to the NLS equation
\beq
iq_{,t'} = q_{,ss} + \frac{1}{2}|q|^{2}q \; \; , \; \;
t'=-\Delta^{2} t \; \; .
\eeq
which describes the motion of a vortex filament in the
localized induction approximation \cite{Hasimoto}\cite{Batch}. We remark that,
in this approximation, the velocity field which governs the motion of the
vortex depends on its curvature $\kappa$ through the relation
\beq
{\bf r}_{,t} = \kappa {\bf b} \; \; ;
\eeq
therefore, since equation (\ref{eq:dNLS2}) has in ${\bf R}^{3}$ a velocity
field
of the same type:
\beq
{\bf r}_{,t} = 2 \Delta \tan(\frac{\varphi_{-1}}{2}) {\bf b}\; \; ,
\eeq
we expect it to be a good candidate for describing the motion of a discrete
vortex in the same approximation.

Consequently, the continuous limit of the linear combination
\beq
q_{,t} = K^{(2)} - 2 K^{(1)}
\eeq
reduces to the complex mKdV equation
\beq
q_{,t'} = q_{,sss} + \frac{3}{2}|q|^{2}q_{,s} \; \; , \; \;
t'= 2\Delta^{3}t \; \; .
\eeq
We remark that, in the degenerate case of the curve on $S^{2}$, $\;
\theta\equiv
0$ and consequently $q,\phi \in {\bf R}$. In this case
we are forced to choose $h_{1}=0$ and only the first hierarchy survives.

{\it Property 3} can also be satisfied through a mechanism (different from that
of equation (\ref{eq:ansatz})) which gives integrable dynamics with sources
(see \cite{DolSan1}\cite{DolSan2}).

\section{Discrete-time dynamics and the Hirota equation}
In this section we consider the discrete curve on $S^{2}(R)$
with constant distance $\Delta = \nu R$ between subsequent points
"moving" in discrete time.

Now we use the following notation: $f$ for $f(k,l), \; k,l \in {\bf Z}$, and
$f_{m}^{n}$ for$f(k+m,l+n)$.
The Frenet equations read
\beq
S_{1} = O_{12}^{\varphi} O_{01}^{\nu} S = A S \; \; .
\eeq
The discrete-time kinematics can be described in terms of angles $\omega$
and $\mu$ which are analogues of the direction and the length of the
velocity field respectively.

\epsffile{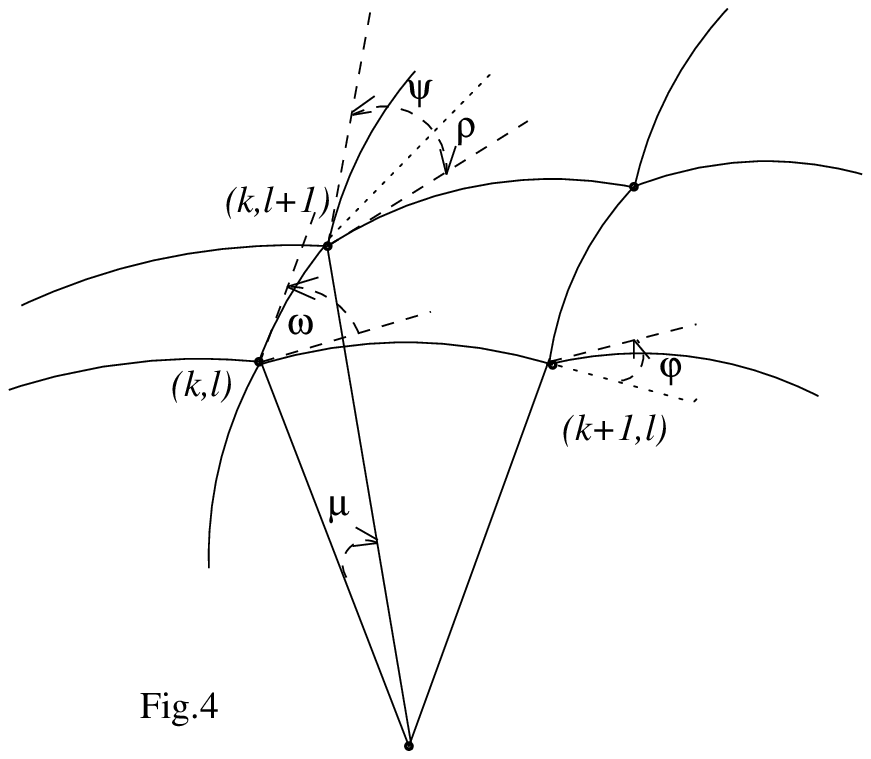}

\noindent The induced kinematics of the Frenet frame reads
\beq
S^{1} = O_{12}^{\rho} O_{01}^{\mu} O_{12}^{\omega} S = B S  \; \; ,
\eeq
where the angle $\rho$ should be calculated from the compatibility
condition
\beq \label{eq:ddcc1}
A^{1} B = B_{1} A \; \; .
\eeq
It turns out that it is convenient to use, instead of $\rho$, another angle
$\psi = \omega^{1} + \rho$, which is the angle of curvature of the
curve $\{ {\bf r}^{n}\}_{n\in {\bf Z}}$.

The compatibility condition (\ref{eq:ddcc1}) written in terms of angles splits
into three equations
\[
\cos \frac{\mu_{1}}{2} \cos\frac{\omega_{1} + \varphi - (\omega_{1}^{1} +
\varphi^{1}  - \psi_{1} )}{2} = \cos \frac{\mu}{2} \cos\frac{\omega -
(\omega^{1} - \psi)}{2} \; \; ,
\]
\beq \label{eq:ddcc2}
\sin \frac{\mu_{1}}{2} \cos\frac{\omega_{1} + \varphi + \omega_{1}^{1} +
\varphi^{1} - \psi_{1} }{2} = \cos \frac{\mu}{2} \cos \frac{\omega +
\omega^{1} - \psi}{2} \; \; ,
\eeq
\[
\sin \frac{\mu_{1}}{2} \sin\frac{\omega_{1} + \varphi + \omega_{1}^{1} +
\varphi^{1} - \psi_{1} }{2} + i
\cos \frac{\mu_{1}}{2} \sin\frac{\omega_{1} + \varphi - (\omega_{1}^{1} +
\varphi^{1} - \psi_{1} )}{2} = \]
\[
{\rm e}^{-i\nu} \left( \sin \frac{\mu}{2}
\sin\frac{\omega + \omega^{1} - \psi}{2} + i \cos \frac{\mu}{2}
\sin\frac{\omega - (\omega^{1} - \psi)}{2} \right) \; \; .
\]
In this paper we consider only the motion subjected to the condition
$\mu\equiv$const. This is (together with the previous condition
$\nu\equiv$const.)  the discrete analog of the Tchebyschev net condition, which
in the continuous case gives the sine-Gordon equation \cite{Tian Chou}.
The first two equations of (\ref{eq:ddcc2}) imply
\beq
\varphi_{-1}^{1}-\varphi_{-1}^{-1} =-( \psi_{1}^{-1}-\psi_{-1}^{-1}) \; \;
\eeq
which asserts the existence of the "potential" $\phi$:
\beq
\varphi = \phi_{2} - \phi \; \; \; , \; \; \;
\psi = - \phi^{2} + \phi \; \; ,
\eeq
and allows to write $\omega$ in terms of it
\beq
\omega = - (\phi^{1} + \phi_{1} ) \; \; .
\eeq
Finally, the third equation of (\ref{eq:ddcc2}) gives the celebrated Hirota
equation \cite{Hir}
\beq
\sin \frac{\phi^{1}+\phi_{1} - \phi_{1}^{1} - \phi}{2} =
\tan \frac{\mu}{2} \tan \frac{\nu}{2}
\sin \frac{\phi^{1}+\phi_{1} + \phi_{1}^{1} + \phi}{2}  \; \; .
\eeq

\end{document}